\documentclass[%
 reprint,
superscriptaddress,
 amsmath,amssymb,
 aps,prper
]{revtex4-2}

\usepackage{graphicx}
\usepackage{dcolumn}
\usepackage{bm}

\usepackage{geometry}
\usepackage{graphicx}
\usepackage{float}
\usepackage{amsmath}
\usepackage{amssymb}
\usepackage{array}
\usepackage{subfiles} 
\usepackage{outlines}
\usepackage{multirow}
\usepackage{longtable}
\usepackage{textgreek}
\usepackage[table,dvipsnames]{xcolor}
\usepackage{booktabs}

\begin{document}

\preprint{APS/123-QED}

\title{Expert covariational reasoning resources in physics graphing tasks}

\author{Charlotte Zimmerman}
\affiliation{Department of Physics, University of Washington}%
\author{Alexis Olsho}
\affiliation{Department of Physics and Meteorology, US Air Force Academy}%
\author{Michael Loverude}
\affiliation{Department of Physics, California State University Fullerton}%

\author{Suzanne White Brahmia}
\affiliation{Department of Physics, University of Washington}%
\date{\today}

\begin{abstract}

Developing and making sense of quantitative models is a core practice of physics. Covariational reasoning---considering how the changes in one quantity affect changes in another, related quantity---is an essential part of modeling quantitatively. Covariational reasoning has been studied widely in mathematics education research, but the language of covariation has only begun to be used in physics education research. We present evidence from 25 individual interviews with physics experts, in which the experts were asked to reason out loud while generating graphical models. We analyze the interviews through the lens of covariational reasoning frameworks from mathematics education research, and determine that the frameworks are useful but do not completely describe the covariational reasoning of the physics experts we interviewed. From our data, we identified reasoning patterns that are not described in the mathematics education research that, together with the mathematics covariational reasoning frameworks, begin to characterize \emph{physics} covariational reasoning.

\end{abstract}

\maketitle

\section{Introduction}
\label{sec:intro}
Learning to model the physical world quantitatively is a key objective of courses in physics, math, and other mathematics-based disciplines \cite{Hestenes1992ModelingWorld, Etkina2005TheInstruction, Brewe2008ModelingPhysics, Hallstrom2019ModelsArgument}. Reasoning quantitatively is at the heart of what it means to ``think like a physicist.''
Some physics education researchers (PER) explore how students engage with mathematics \textit{conceptually} when they reason about physics, \cite{Uhden2012ModellingEducation, Karam2014FramingElectromagnetism, Korff, Caballero2015UnpackingHere, Ibrahim2017HowPerformance}, and claim that quantitative reasoning has some different characteristics in a physics setting than in a purely mathematical one \cite{Sherin2001HowEquations, Bing2007, Kuo2013HowProblems, Hu2013, Redish2015LanguageEpistemology}. 

Foundational to physics reasoning is relating the change in one physical quantity to the change in another. Physics experts notice how a quantity changes, as well as its rate of change. For example, an expert considering an RC circuit might think first about the immediate moments after a switch is closed---that the voltage across the capacitor increases quickly with time---and then that the rate of increase diminishes until the voltage reaches a steady value equal to the battery voltage. This kind of reasoning is described in mathematics education research as \emph{covariational reasoning}.

\begin{figure*}
    \centering
    \includegraphics[width=0.85\textwidth]{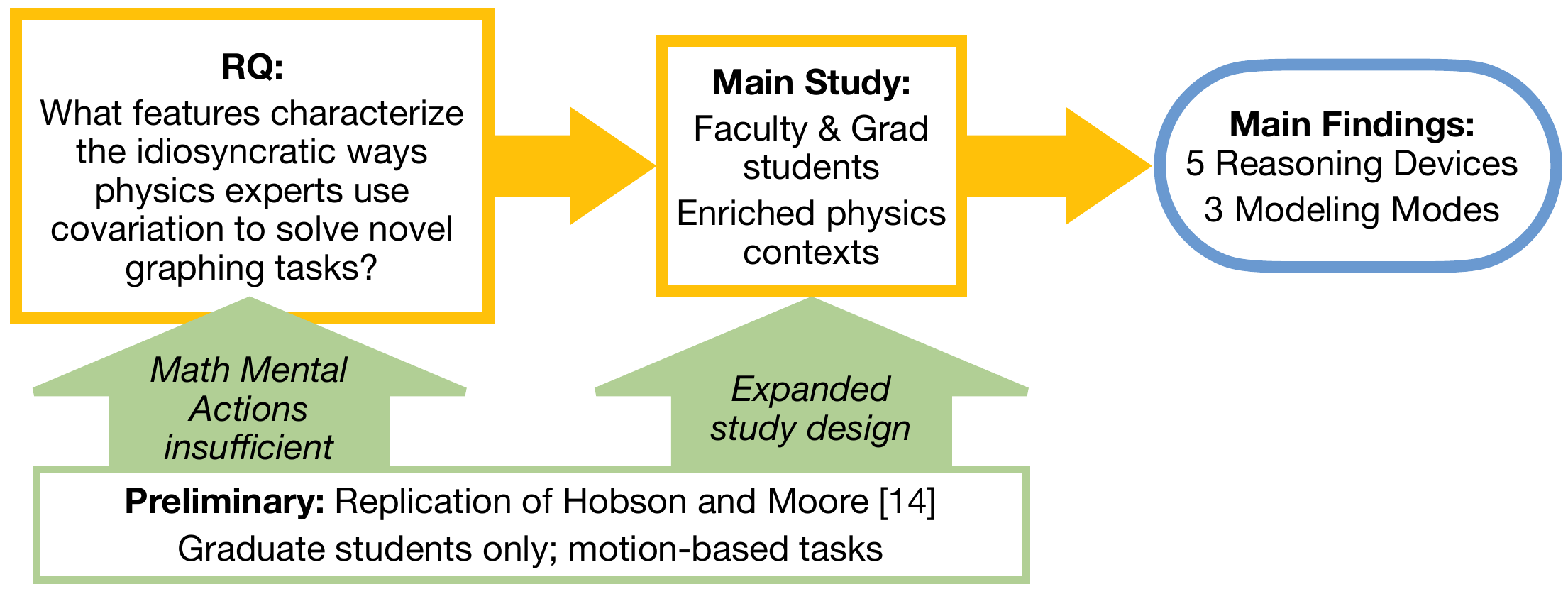}
    \caption{A diagram of the design and conceptualization of the study. The results of the replication experiment informed the research question and design of the study we present in this paper.}
    \label{fig:timeline}
\end{figure*}

This paper contributes to the body of evidence that quantitative reasoning in physics contexts has different features than in purely mathematical ones. Here, we report on findings from experiments conducted to help characterize some of features of covariational reasoning in physics, in the context of graphing tasks. 

There is little published in PER that uses the language of covariational reasoning, as its use originates in mathematics education research. Therefore, we turn to the mathematics education research literature for guidance. Conceptualizing the mathematical functions that characterize common covariational relationships seen in physics is important in pre-calculus and calculus instruction, as well as physics \cite{Thompson1994ImagesCalculus, Carlson2002ApplyingStudy, Ojose2011mathematicsUse, Johnson2015TogetherChange, Thompson2017, Hobson2017, Jones2022MultivariationReasoning}. A framework of ``mental actions'' that operationalizes covariational reasoning in mathematics contexts has been developed and updated over the past two decades, describing specific ways in which this reasoning can be observed in novices and experts \cite{Carlson2002ApplyingStudy, Thompson2017, Jones2022MultivariationReasoning}. This framework from mathematics education research provides a foundation for the research described in this paper.

To validate the use of this framework to analyze physics reasoning, we first replicated an experiment that was originally done with expert mathematicians by researchers Hobson and Moore \cite{Hobson2017},  using physics experts as the subjects of the study \cite{Zimmerman2019TowardsPhysics}. Consistent with prior work in physics education, we found that---while the mathematics frameworks overlap in important ways with physics reasoning---they do not capture some important features of physics expert reasoning. 

Having confirmed both that the frameworks are useful in physics contexts, and that there is a void in the research literature, we sought to produce a preliminary characterization of physics expert covariational reasoning. The study and the findings we report on in this paper address the research question, ``What features characterize the ways physics experts use covariation to solve novel graphing tasks?'' After validating the utility and limitations of the existing mathematics frameworks, we conducted a study that was modeled on the Hobson and Moore experimental design, but included a broader group of experts and tasks that both spanned the expert space, and more richly involved physics contexts. Figure~\ref{fig:timeline} shows a simplified description of our conceptualization of the research study.

Five ``reasoning devices'' and three ``modeling modes'' emerged from our study as characteristics of physics covariational reasoning in the context of graphing tasks. The ``reasoning devices'' are a set of ways physics experts were reasoning that are related to covariational reasoning, but are distinct from the mathematics mental actions (MMA) in existing frameworks. The ``modeling modes'' are larger patterns of physics expert reasoning that rely on both the reasoning devices and the MMA. We suggest that these modeling modes describe a broad set of thought processes related to physics covariational reasoning. However, as our data are limited to graphing contexts, we do not claim the reasoning devices or modeling modes span the space of all physics covariational reasoning. Therefore, we present our results as a hypothesis that contributes to developing a broader description of physics covariational reasoning. 

\section{Background}
\label{sec:background}

\renewcommand{\arraystretch}{1.5}
\setlength{\tabcolsep}{6pt}
\begin{table*}[]
    \centering
    \begin{tabular}{p{0.07\textwidth} p{0.22\textwidth} p{0.25\textwidth} p{0.35\textwidth}}
        \toprule
         Label & Mental Action \cite{Carlson2002ApplyingStudy, Thompson2017} & Brief Description \cite{Jones2022MultivariationReasoning} & Example Behavior \\
         \toprule
         MMA~1 & Recognize Dependence & Identify variables that are dependent & Labeling axes \\
         MMA~1.5 & Precoordination & Asynchronous changes in variables & Articulating that first, one quantity changes, and then the other changes \\
         MMA~2 & Gross Coordination & General increase/decrease relationship & Describing that as one quantity increases, another decreases \\
         MMA~3 & Coordination of Values & Tracking variable's values & Plotting points \\
         MMA~4 & Chunky Continuous & Values changing in discrete chunks & Articulating that as one quantity doubles, the other triples \\
         MMA~5 & Smooth Continuous & Continuous, simultaneous changes & Describing that the quantities vary together, smoothly and continuously\\
         \bottomrule
    \end{tabular}
    \caption{A summary of the covariational reasoning mental actions (MMA) used in this study. The framework we provide here is a slightly modified version of that summarized by Jones in 2022 \cite{Carlson2002ApplyingStudy, Thompson2017, Jones2022MultivariationReasoning}. We use names and descriptions from Jones' summary. Numeric labels and example behaviors were added by the authors and correspond to the original mental actions from the 2002 framework, with the addition of 1.5 for ``precoordination,'' for ease of reference within this paper.}
    \label{tab:StevensSummary}
\end{table*}

In this section, we situate our work in the current literature on covariational reasoning in mathematics and physics education research. An essential part of covariational reasoning is the mathematical formalism used to represent the relationships between quantities. Therefore, we begin by presenting our perspective on reasoning about quantities as compared to reasoning about acontextual variables. We then discuss covariational reasoning as defined by mathematics education researchers, and examine work in physics education research on reasoning about rates of change through a covariational reasoning lens. We provide a brief summary of expert mathematics covariational reasoning as described by Hobson and Moore \cite{Hobson2017}, and then finally summarize the results of our preliminary experiment replicating the work of Hobson and Moore. This preliminary work forms the groundwork of the research presented in this paper (Fig.~\ref{fig:timeline}) \cite{Zimmerman2019TowardsPhysics, Zimmerman2020ExploringPhysics}.

\subsection{Blending and Covariational Reasoning}
Reasoning about quantity takes many forms; prior research has described the importance of multiple representations in developing quantitative models, including using equations, graphs, and diagrams to guide reasoning \cite{Hestenes1992ModelingWorld, Etkina2006TheInstruction, Brewe2008ModelingPhysics, Treagust2017MultipleEducation, Rolfes2022Mono-Thinking}. Essential to all forms of modeling, however, is making sense of the meaning of the quantities involved, regardless of how they are represented. Recent work suggests that reasoning mathematically in physics can be viewed as a blend of conceptual mathematical ideas and reasoning about the physical meaning of the quantities \cite{Bing2007, Hu2013,  Redish2015LanguageEpistemology, Taylor2018SoKinematics, Huynh2018BlendingNegative-ness, Eichenlaub2019, WhiteBrahmia2019QuantificationPhysics, VanDenEynde2020}. Schermerhorn and Thompson define \emph{symbolic blending}, which describes how the meaning of physical quantities is interwoven with the meaning of the mathematical symbols used to represent them \cite{Schermerhorn2023MakingBlending}. We take the blended nature of physics and math as a foundation of our research perspective. Therefore, we suggest that \emph{physics} covariational reasoning is necessarily grounded in the meaning of the relevant physical quantities.

In mathematics education research, covariational reasoning is defined broadly as reasoning about how two or more quantities vary with respect to one another \cite{Thompson2017}. It has been studied in the context of mathematics for several decades, and is an essential tool for reasoning about a relationship between two or more quantities \cite{Thompson1994ImagesCalculus, Confrey1995, Saldanha1998Re-thinkingVariation, Carlson2002ApplyingStudy, Moore2013, Castillo-Garsow2013, Johnson2015TogetherChange, Hobson2017, Thompson2017, Paoletti2018, Ely2019, BYERLEY2019100694}. Mathematics education researchers Carlson et al. developed a framework of \emph{mental actions} and associated behaviors that operationalizes covariational reasoning, which has been revised several times based on continued research \cite{Carlson2002ApplyingStudy, Thompson2017, Jones2022MultivariationReasoning}.  We find Jones' simplified and streamlined framework is most productive for our analysis \cite{Jones2022MultivariationReasoning}, and a slightly modified version is shown in Table~\ref{tab:StevensSummary}. In this paper, we refer to these as the mathematics mental actions (MMA).

Physics education researchers often describe student reasoning about related changes using the language of proportional reasoning \cite{Boudreaux2015, HankyPanky, maloney2010NTIPERs:Mechanics, Sherin2001HowEquations} and scaling \cite{Arons1976CultivatingCourse, Trowbridge1981InvestigationDimension, Bissell2022IllustratingLaw}. In the context of covariational reasoning, we argue that proportional reasoning is a \emph{linear version} of covariation and that scaling 
aligns with the fourth MMA: Chunky Continuous, in which one might ask, ``if I double this, what happens to that?'' (see Table~\ref{tab:StevensSummary}). 

In recent PER studies that focus on mathematical modeling in physics, covariation---as defined in mathematics education---offers  valuable explanatory power in characterizing reasoning \cite{Taylor2018SoKinematics, Emigh2019, Olsho2022CharacterizingModeling, may2021StudentsStudy, AkinyemiSolutionStudents}. In science education research more broadly, covariation has been studied in the context of single and multiple representations \cite{Rolfes2022Mono-Thinking,Jimenez2024MathematicalStudy,Altindis2024ExploringInterpretation}. We suggest physics and science education researchers find the language valuable  because it allows for descriptions of reasoning about change across a wide variety of continuous, functional relationships between two or more quantities \cite{Jones2022MultivariationReasoning}.

A common feature of covariational reasoning tasks is graphical representations. This study is framed around asking experts to generate graphical representations from animations. There is a long tradition of examining multiple representations as a framework for learning in physics education research (for a review of this work, please refer to \cite{Rosengrant2007AnRepresentations}. Due to the expert level of our study subjects, we assume their expertise in interpreting graphs and moving fluidly between representations. Therefore, we choose to take a more holistic approach towards being able to identify the kinds of covariational reasoning they use when making sense of both contexts, rather than focus on their translation between representations.

\subsection{Expert mathematics Covariational Reasoning: Hobson and Moore's 2017 Study}
\label{sec:backgroundHandM}
\begin{figure}
    \centering
    \includegraphics[width=0.45\textwidth]{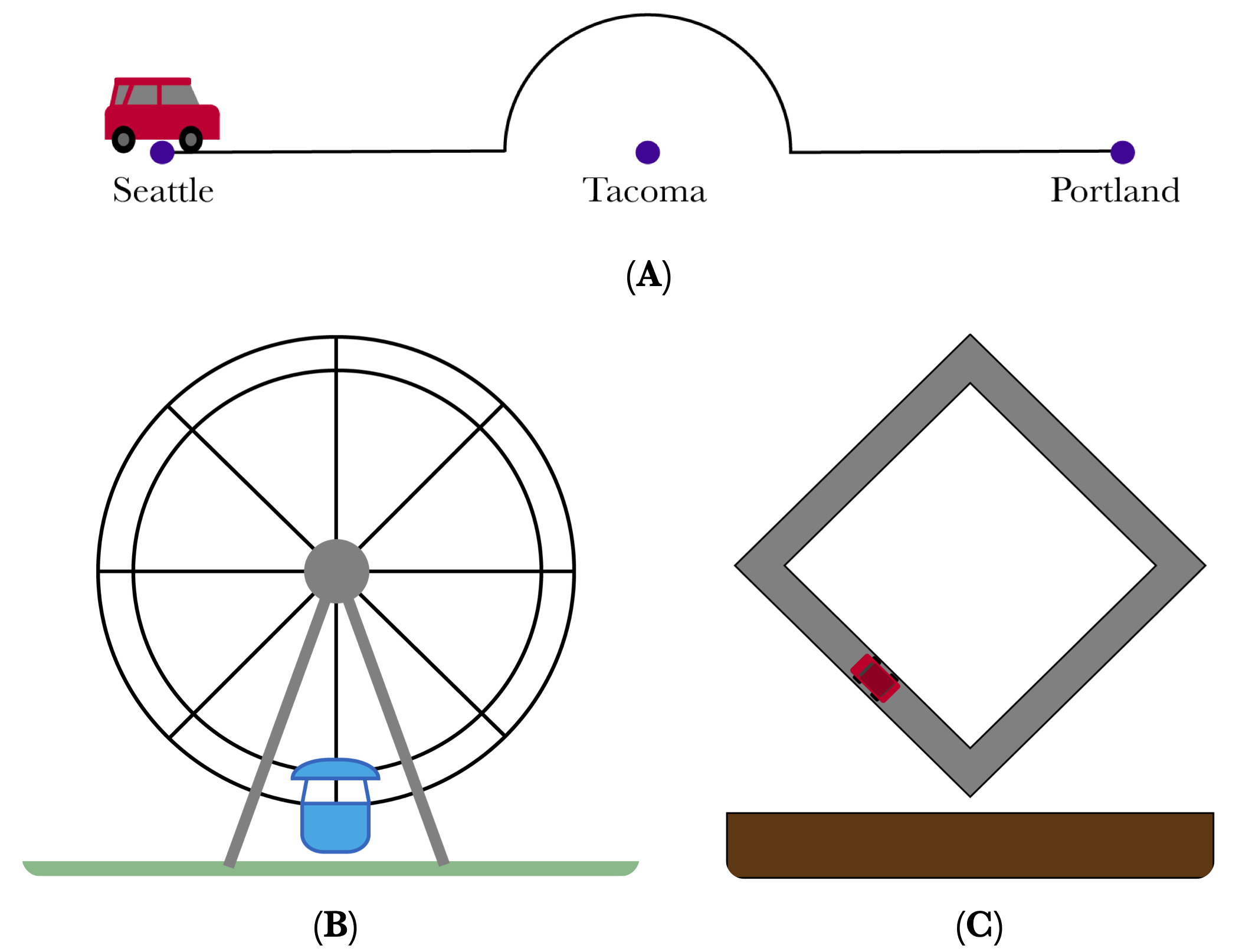}
    \caption{Stills from the animations used in the preliminary experiment: (A) Going Around Tacoma, (B) Ferris Wheel, and (C) Square Track. All tasks asked participants to create a graph that related a particular quantity (the distance from the car to Tacoma, the height of the cart, and the distance from the cart to the wall, respectively) to the total distance traveled of an object in constant motion. These tasks are identical to those in the original study conducted by Hobson and Moore \cite{Hobson2017}, with the exception that we changed the place names in Going Around Tacoma.}
    \label{fig:firstround}
\end{figure}

Mathematics education researchers Hobson and Moore examined how expert mathematicians use covariational reasoning by asking several mathematics graduate students to complete novel graphing tasks \cite{Hobson2017}. Hobson and Moore considered graduate students to represent experts in the context of the introductory calculus problems used in the study, as graduate students are expected to ``have sufficient experience engaging in tasks involving representing and reasoning about quantities'' \cite{Hobson2017}. The graduate students were shown animations developed by the researchers for three tasks identical to those shown in Fig.~\ref{fig:firstround}, with place names adjusted to the local region. 

All three tasks asked participants create a graph to represent the relationship between between a particular quantity (e.g., the height of a Ferris wheel cart) and the total distance traveled of an object moving at constant speed. Relating two distances was an intentional choice by the researchers, who noted the facility with time-based motion tasks that experts in math-based disciplines typically acquire. The research found that mathematics graduate students were likely to:
\begin{itemize}
    \item[(1)] Compare the change of the quantities in the prompt directly to one another (in some cases, with prompting from the interviewer), and
    \item[(2)] Divide the domain of the task, or large sections of the motion, into small, equally-sized sections to compare the change in the two quantities and then map these equal segments to their graphs \cite{Hobson2017}.
\end{itemize}
We were interested in what similar and different behaviors would emerge if we gave the same tasks to physics graduate students.

\subsection{Preliminary Experiment}
\label{sec:backgroundprelim}
An early version of this work reported that there are elements of physics covariational reasoning that are distinct from those in mathematics \cite{Zimmerman2019TowardsPhysics, Zimmerman2020ExploringPhysics}. In particular, physics graduate students were found to:
\begin{itemize}
    \item[(1)] Use other quantities than those given in the task (i.e. time in place of distance), and
    \item[(2)] Create a graph by applying a model they already knew, or considering the rate of change only near a small number of physically meaningful points.
\end{itemize}
We consider these behaviors distinct from the results reported by Hobson and Moore, and take this as evidence that further investigation into characterizing \emph{physics} covariational reasoning is warranted. 

An equally important finding not discussed in our earlier published work is that some MMA from Table~\ref{tab:StevensSummary} emerged from the initial coding scheme in the preliminary study, and are useful toward describing physics covariational reasoning. While we consciously chose an analysis framework that did not use the MMA in our codebook a-priori to ensure the results emerged from the data directly, we nevertheless found that MMA~1~4 were all present in the final codebook. (See \cite{Zimmerman2019TowardsPhysics} for more details on the study itself.) However, we were unable to relate any of the code categories that emerged from our analysis to MMA~5 (Smooth Continuous)---none of the physics graduate students were observed to reason about the second derivative to the extent described in the mathematics literature. The full codebook, analyzed through the lens of existing mathematics frameworks, is available in the Appendix.

We recognize all of these findings as important results, and they motivate the work presented in this paper. However, the preliminary experiment was limited in scope. The data are not sufficient to characterize the distinct features of reasoning physics graduate students used in addition to the MMA. The graduate students were so familiar with the motion-based contexts that few related the two distances in the tasks directly (as observed by Hobson and Moore); they used time and familiar models of constant motion instead. Finally, this experiment only included graduate students and therefore does not represent a complete picture of expertise.

\section{Methods}
\label{sec:mainMethods}

\renewcommand{\arraystretch}{1.2}
\begin{table*}[]
    \centering
    \begin{tabular}{p{2cm} p{4cm} p{6cm}}
        \toprule
         &  2019 Study & This Study\\
         \toprule
         Subjects & 10 Graduate Students & 5 Graduate Students from 2019 Study\\
         & & 5 New Graduate Students \\
         & & 5 Physics Faculty \\
         \hline
         Tasks & Going Around Tacoma & Gravitation / Electric Charge \\
         & Ferris Wheel & Ferris Wheel \\
         & Square Track & \\
         & & Drone \\
         & & Intensity \\
         \bottomrule
    \end{tabular}
    \caption{A summary of the differences between the populations and tasks between our previous, preliminary study in 2019 and the study we report on here.}
    \label{tab:summary}
\end{table*}

This research was conducted at a large, R1 university in the Pacific Northwest. The population includes five tenure-track or tenured faculty members and ten physics graduate students. To account for whether experience with the prior tasks had an effect, we included five graduate students from the preliminary experiment and recruited five new graduate students. An effort was made to ensure the study population was representative of a variety of physics sub-fields, and was diverse across race, country of origin, and gender. 

The study consisted of a series of one-on-one, think-aloud interviews, in which the participant was asked to complete three graphing tasks. Each task included a written prompt and an animation, and asked the participant to create a graph to represent the relationship between two quantities specified in the prompt. The interviewer followed up with questions after the participant expressed they were finished with all of the tasks.  The interviews were audio recorded and the participants' written work was collected; initial transcripts were created automatically using the Otter.ai software program \cite{2019Otter} and were subsequently hand corrected. One member of the research team conducted all of the interviews, and the same team member corrected all of the transcripts.

\begin{figure}
    \centering
    \includegraphics[width=0.45\textwidth]{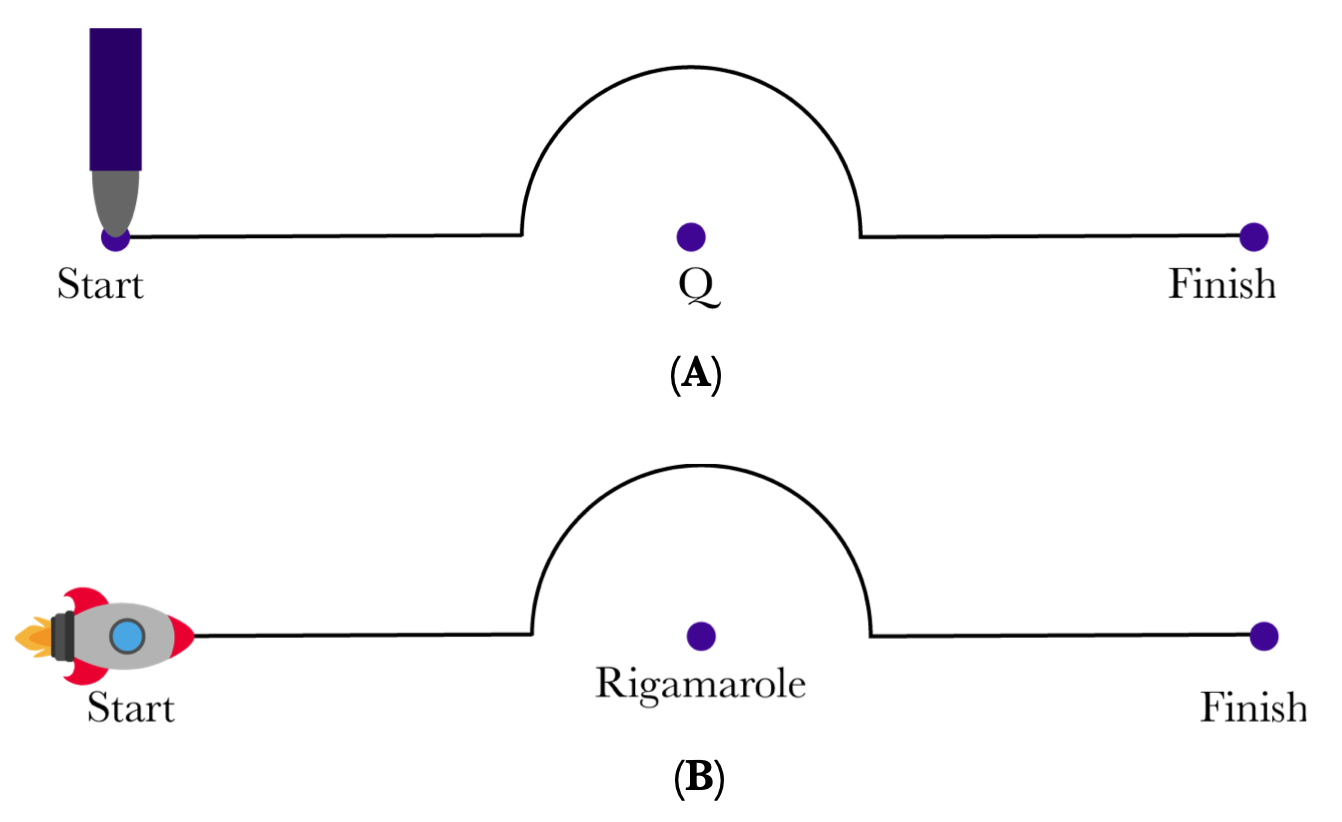}
    \caption{Stills from the animations associated with the first task administered in the main study. Participants either saw (A) Electric Charge or (B) Gravitation. The tasks prompt participants to create a graph that relates either the electric or gravitational potential and the total distance traveled of the probe or spaceship, as it moves (under either a guiding hand or propulsion) at constant speed from start to finish.}
    \label{fig:potentialtasks}
\end{figure}

\begin{figure}
    \centering
    \includegraphics[width=0.45\textwidth]{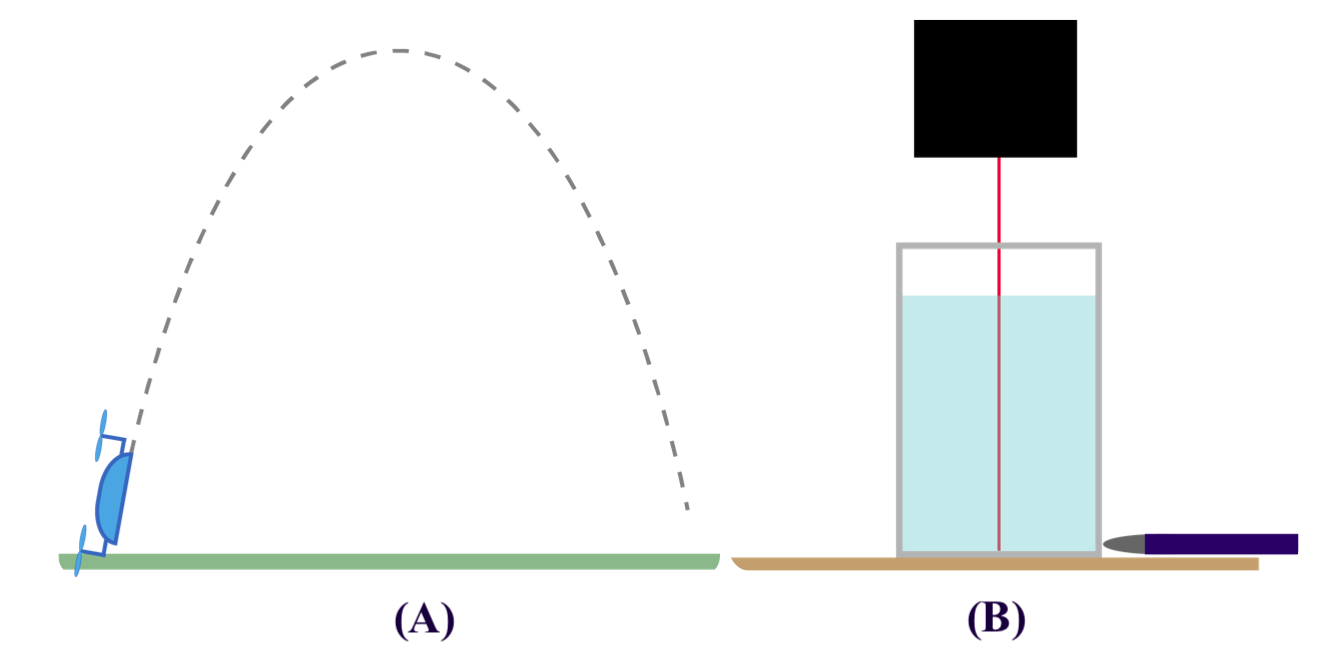}
    \caption{Stills from the animations associated with the second and third tasks administered in Study 2: (A) Drone, in which participants were prompted to create a graph that relates the height of the drone to the angle of the drone, flying in the arc shown; and (B) Intensity, in which participants were prompted to create a graph that relates the intensity measured by the probe at the side of the liquid column to the total distance traveled of the probe as it moves from the bottom of the column to the top.}
    \label{fig:droneandlight}
\end{figure}

The tasks can be found in Figures~\ref{fig:potentialtasks} and~\ref{fig:droneandlight}. We designed two isomorphic versions of the first task from the preliminary experiment, Going Around Tacoma (Fig.~\ref{fig:firstround}A). One version, Electric Charge, asks the participants to relate the electric potential measured by a probe and the total distance traveled by the probe as the probe moves around a charge, $Q$ (Fig.~\ref{fig:potentialtasks}A).  The other version, Gravitation, asks participants to relate the gravitational potential of the rocket ship-planet system and the total distance traveled by a rocket ship as it moves around a planet called Rigamarole (Fig.~\ref{fig:potentialtasks}B). Half of the graduate students were given Electric Charge, and the other half were given Gravitation. We found there was little difference in covariational reasoning, but that the graduate students were more comfortable with Gravitation. As covariation is the focus of this study, we therefore chose to reduce cognitive load by giving Gravitation to all the faculty participants. 

Drone (Fig.~\ref{fig:droneandlight}A) was designed to relate two familiar quantities---height and angle---in a novel way. It asks participants to relate the height of a drone to the angle the drone's base makes with the horizontal axis, with the constraint that the drone is designed to tilt such that its base is always aligned with its velocity vector. Finally, Intensity (Fig.~\ref{fig:droneandlight}B) depicts a laser shining through a column of liquid and asks participants to relate the intensity of light measured by a probe and the total distance traveled by the probe as it moves upward along the column of liquid. All of the participants saw both Drone and Intensity. 

We also gave the faculty participants the Ferris Wheel task (Fig.~\ref{fig:firstround}B) as it was the most fruitful of the preliminary experiment tasks. Rather than re-administer Ferris Wheel to graduate students, we included the data from the preliminary experiment concerning Ferris Wheel in this data corpus. These data were re-analyzed alongside the rest of the interviews as part of this study.

We recognize these tasks are vague in many ways: the charge of $Q$ is not explicitly noted in Electric Charge, the angle at which the Drone flies is difficult to quantify from the animation, and the absorption of the liquid in Intensity is not provided. We consider the lack of specific information a feature of this study; the goal is to learn about what physics experts do when confronted with less familiar contexts. We are therefore not interested in the correctness of the answers. Rather, we seek to learn about the kinds of reasoning the experts used, in the context of the assumptions they made.

\begin{figure}
    \centering
    \includegraphics[width=0.45\textwidth]{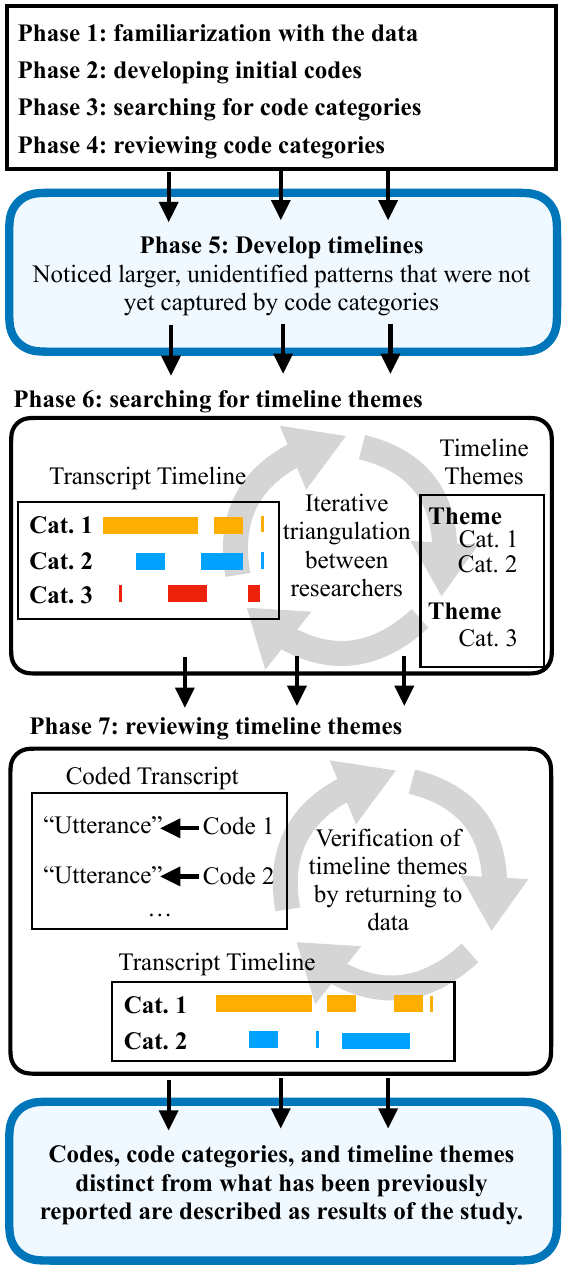}
    \caption{An outline of our analysis methods.}
    \label{fig:methodsmain}
\end{figure}

To analyze the data, we chose a thematic analysis framework \cite{Nowell2017ThematicCriteria} for two reasons: (1) the goal of the study is to seek broad patterns using multiple approaches, and (2) thematic analysis ensures reliability through researcher triangulation and discussion which more accurately represents the results of our hypothesis generating experiment \cite{Guest2011AppliedAnalysis, Braun2006UsingPsychology, Richards2018AAnalysis}. This approach is consistent with recent recommendations for qualitative research in PER \cite{Barth-Cohen2023MethodsResources}. The phases of our analysis were:
\begin{enumerate}
    \item Familiarization with the data;
    \item Initial coding, with iterative researcher triangulation;
    \item Searching for themes across the codes, using a card sorting task (results in code categories) \cite{Chi1981CategorizationNovices, Schoenfeld1982ProblemSolvers};
    \item Reviewing code categories, through iterative researcher triangulation;
    \item Development of timeline charts to probe larger-scale patterns across code categories;
    \item Searching for patterns in timeline charts across code categories; and
    \item Reviewing patterns in timeline charts, through iterative triangulation and validation with the original text.
\end{enumerate}
We will describe these phases in detail in the following paragraphs; a summary can be found in Fig.~\ref{fig:methodsmain}.

\subsection{Phases 1-4}
Phase 1 was addressed by the lead researcher taking both the role of interviewer and lead analyst. In phase 2, three members of the four person research team met to discuss one representative transcript and generate an initial codebook using Process and In Vivo coding \cite{Saldana2016TheResearchers}. Based on the results of the preliminary experiment, we also included the MMA in Table~\ref{tab:StevensSummary} a-priori. Two of these three members then individually coded a separate transcript using the initial codebook. This first round of triangulation had 74\% agreement between codes; discrepancies were resolved through discussion and the result was an updated version of the initial codebook. 

The lead researcher used this updated version to code the remaining transcripts. As an additional level of triangulation, the fourth member, who was not involved in the original code development, coded another transcript. The fourth member and the lead coder began this coding exercise with 76\% agreement; disagreements were resolved and resulted in a finalized codebook after discussion amongst the entire research team. The lead researcher used this final codebook to code the remaining data.

In phase 3, two members of the research team separately engaged in a card-sorting task in which they categorized coded sections of the transcript into larger groups based on the ways the participants expressed their reasoning (for more information about card-sorting tasks, and their uses in qualitative research, see Refs. \cite{Chi1981CategorizationNovices, Schoenfeld1982ProblemSolvers}). They resolved inconsistencies through discussion. The resulting groups of codes are called ``code categories.'' Finally, in phase 4, the data were binned according to the code categories and the categories were verified by comparing back to the original transcripts across the entire set of interviews. Phase 4 resulted in the code categories defined in Table~\ref{tab:maincategories}. The full, finalized codebook can be found in the Appendix.

We include our analysis details to emphasize that the codebook and code categories emerged out of several, iterative discussions and revisions until all members agreed they were representative of the data. As a result, we do not report inter-rater reliability statistics because they do not hold much meaning in a context where the researchers were in continuous discussion \cite{Cheung2023TheEducation, Barth-Cohen2023MethodsResources, Richards2018AAnalysis}. We intend our results to be interpreted as a hypothesis to be tested. It is not our expectation that another researcher might use this methodology and develop the same codebook, but instead that an expert in mathematical reasoning and physics education research might recognize the patterns we share in these data and view our results as one possible interpretation of the interviews.

\begin{table*}
    \centering
    \begin{tabular}{p{3.5 cm} p{10.5 cm}}
        \toprule
        Code Category & Definition \\
        \hline
        Connection to the Task & Connecting a representation to the physical meaning described in the task.\\
        Quantity & Recognizing, defining, and symbolizing relevant physical quantities. \\
        Function & Naming and assigning mathematical functions to relate two quantities. \\
        Mental Actions & MMA~1-4 defined by mathematics education research. \\ 
        Graphing & Behaviors associated with drawing a graph. \\
        \bottomrule
    \end{tabular}
    \caption{The relevant code categories that resulted from phases 1-4 of the analysis in the main study, and their definitions. For more details, see the full codebook in the Appendix.}
    \label{tab:maincategories}
\end{table*}
\begin{figure*}
    \centering
    \includegraphics[width=\textwidth]{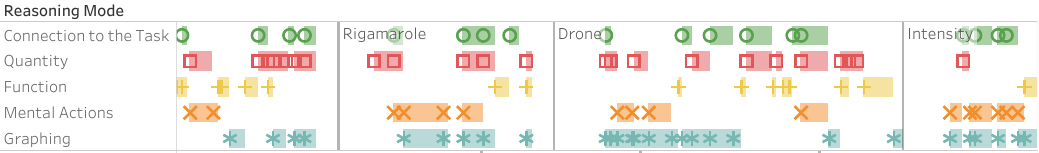}
    \caption{An example timeline chart from the main study. The horizontal axis represents the fraction of time on task. The code categories are represented on the vertical axis. The symbol represents the time that category was coded, and the band represents the duration that the interviewee was observed to be engaged with that kind of reasoning. The code categories are defined in Table~\ref{tab:maincategories}.}
    \label{fig:ganttChartExample}
\end{figure*}

We did not see any significant discrepancies between graduate students who participated in the previous study and those that had not during coding nor during the card sorting task. We interpret this to mean that prior participation has a minimal effect, and therefore consider the dataset as a whole for the remainder of the analysis presented here.
\subsection{Phases 5-7}
Phases 5 through 7 make use of timeline charts. Timeline charts are plots that show the code categories assigned to a participant over the time of their interview. Each plot shows the fraction of total time on task along the horizontal axis, and a mark for each code category at the each moment it was assigned (an example is shown in Fig.~\ref{fig:ganttChartExample}, and the code categories from phase 4 are defined in Table~\ref{tab:maincategories}). The tails associated with each mark show for how long the participant was engaged in reasoning represented by that code category. This method allowed the research team to find patterns between categories in the diagrams and then draw conclusions about how code categories may be interacting, grounding the analysis in the data. 

In phase 6, the research team sought patterns in the code categories by looking for categories that occur at the same time and categories that rarely appear together. For example, in Fig.~\ref{fig:ganttChartExample}, ``Quantity'' often appears at the same time as the code category ``Connection to the Task''. In contrast, ``Function'' often appears before or after---but rarely at the same time as---the code category ``Mathematics Mental Actions''. 

To review and verify the patterns we characterized from the timeline charts (phase 7), the lead researcher went back into the transcripts to examine how the participants described their thought processes and what behaviors were observed by the interviewer at those moments where code categories coincided or were apart. Over the course of several research team discussions, the team went iteratively back and forth between the timeline charts and the transcripts to characterize the patterns we observed. 

During phase 7, some new patterns emerged that we had not identified in phase 6, some patterns identified in phase 6 were revised and clarified, and others patterns from phase 6 were set aside. For example, in Fig.~\ref{fig:ganttChartExample}, the code category ``Graphing'' often appears at the same time as ``Mathematics Mental Actions'', but not every time. The transcript reveals that this is because the MMA describe reasoning that is closely tied to graphical behaviors. For example, identifying that two quantities are related is tightly associated with the behavior of labeling the axes of a graph. The relationship between MMA and graphing behaviors is well documented in mathematics education research. Therefore, we did not further characterize this relationship.

\section{Findings}
From the codebook and timeline chart analysis, three main patterns emerged: 
\begin{enumerate}
    \item[0.] The mathematics covariational reasoning framework is productive for describing some of the covariational reasoning used by physics experts.
    \item[1.] Physics experts also used a set of five distinct ``reasoning devices'', distinct from reasoning characterized by existing frameworks.
    \item[2.] Physics experts broadly engaged in three patterns of reasoning, or ``modeling modes'', when they were solving graphing tasks.
\end{enumerate}
We refer to the first result as the ``zeroth'' result because it reproduces the result from the preliminary experiment. In particular, every participant was shown to use MMA~1-4 and we did not see any evidence of MMA~5. The wide applicability of these codes demonstrates the MMA are productive for describing the covariational reasoning used by physics experts solving novel graphing tasks. A more complete analysis can be seen in the full timeline charts, available in the Appendix. Results 1 and 2 are further explorations of how physics experts covaried, towards characterizing \emph{physics} covariational reasoning.

Our research question is, ``What features characterize the ways physics experts use covariation to solve novel graphing tasks?'' To characterize these ``other features of reasoning'', we (1) looked carefully at our codebook for individual codes that require or rely on reasoning about change and rates of change, and (2) examined the timeline patterns between code categories that included the MMA code category. The first approach resulted in five ``reasoning devices'': features of reasoning that involve covariation but are not well described by the MMA. The second resulted in three ``modeling modes'': larger patterns of problem solving that experts engaged in, that include the use of both the MMA and the reasoning devices. We suggest that together, the MMA, reasoning devices, and larger modeling modes are representative of covariational reasoning in physics. In this section, we will define the reasoning devices and modeling modes that we identified and present an emerging hypothesis that contributes to describing physics covariational reasoning.

\subsection{Reasoning Devices}
\label{sec:resultsRD}

\renewcommand{\arraystretch}{1.5}
\setlength{\tabcolsep}{6pt}
\begin{table*}[]
    \centering
    \begin{tabular}{
        p{0.2\textwidth}  p{0.32\textwidth} p{0.4\textwidth}}
    \toprule
        Reasoning Device 
            & Description & Example \\
    \toprule
        Compiled Models
            & The strong association between a particular functional form and a physical context.
            & ``And as you get closer the potential is one over R. So I'm literally just plotting one over R.'' \\
        Proxy Quantity 
            & A quantity used to replace another, often to simplify the task.
            & ``It's going at a constant rate. So time maps one-to-one to, uh, distance traveled.''\\
        Regions of Consistent Behavior
            & Separating a domain into sections that are modeled by the same function.
            & ``When they’re doing that loop around Tacoma---looks like half a circle. So that’s going to be a constant distance.''\\
        Physically Significant Points
            & Plotting only those points that hold physical significance.
            & ``So it's starting and finishing the same distance away\ldots[plots end points]''\\
        Neighborhood Analysis
            & Examining the rate of change of one quantity with respect to another over a ``small chunk'' around a physically significant point.
            & ``So it's, it's angular velocity has to be faster when it's in the narrow part of the peak\ldots[and] I have more, more, change in the angle for smaller change in the height.'' \\
    \bottomrule
    \end{tabular}
    \caption{The reasoning devices identified from the physics experts' descriptions of their reasoning as they thought out loud while solving novel graphing tasks.}
    \label{tab:RD}
\end{table*}

We identified five predominant, individual codes that are connected to thinking about how changes in one quantity affect changes in another, yet are distinct from what is described in existing frameworks of covariational reasoning (see the Appendix for a full analysis). We call these ``reasoning devices,'' because the codes are a result of a combination of how the experts described their own reasoning while talking out loud and the graphing behaviors we observed the participants perform. 
In the following subsections, we describe each of the reasoning devices we identified. A summary of the reasoning devices can be found in Table~\ref{tab:RD}. \\

\noindent
\textbf{Compiled Models} \\
We define the use of ``compiled models'' as applying an already-known model to a particular physical context. This occurs when the physical context is already ``compiled'' in one's mind with a common functional relationship between two or more quantities. The ``$1/r^2$'' model for conservative forces is an example familiar to all physicists. Having a compiled model allows the reasoner to use wide variety of resources about how two quantities relate to one another. We observed experts use compiled models frequently for tasks that had familiar physical contexts. Compiled models involves the application of a model and often does not consistently engage MMA beyond a trivial use of MMA~1, but it was such a useful tool for the experts that we consider it an essential part of reasoning about how two quantities are related. All experts were observed to use a variety of compiled models.

For example, while puzzling about the Intensity task, one graduate student stated:
\begin{center}
\begin{minipage}[c]{0.4\textwidth}
``This is going to depend on how much light gets scattered by the water. And\ldots I'm tempted to think that this is going to be an exponential decay\ldots because the scattering is a probabilistic process.''
\end{minipage}
\end{center}
This graduate student then quickly generated a graph using an assumption about the initial intensity, and drawing an exponential curve. The evidence of compiled models across all experts and tasks validated our initial observations in the preliminary experiment that experts used and applied previously known models to make sense of the tasks.

We observed a number of compiled models in various contexts: a linear function that relates distance and time, $d \propto t$, in  constant motion contexts;  an exponential function that relates intensity and depth, $I \propto e^{-y/y_0}$, in scattering light contexts;  and an inverse function that relates potential energy and distance from the source, $U \propto 1/r$, in potential energy contexts. 

These associations manifested verbally (``So potential goes like $1/R$''), symbolically, and graphically, suggesting that the models hold deep meaning to the user. 
Because compiled models rely on familiar mathematical models, they are one way that expert physicists were able to arrive at an answer to the tasks without explicitly analyzing how the change in one quantity affects the change in another. \\

\noindent
\textbf{Proxy Quantities} \\
We define a ``proxy quantity'' as a quantity that is not present in the prompt of the task, but brought in by the interviewee to replace another. We observed that it was often done to make the task easier to think about. The use of proxy quantities implicitly relies on MMA~1, as the proxy quantity is related to the quantity it replaces. All physics experts were observed to use proxy quantities.

During our interviews, we found use of proxy quantity occurred most often when participants substituted time for total distance during constant motion tasks. For example, a graduate student working on Ferris Wheel said:
\begin{center}
\begin{minipage}[c]{0.4\textwidth}
``It's moving at constant speed, so that’s [the distance is] just going to map to time.''
\end{minipage}
\end{center}
\noindent
They then proceeded to solve Ferris Wheel by thinking out loud about time instead of total distance while simultaneously labelling the horizontal axis as total distance.

We observed that proxy quantities were also used to relate quantities that were not motion-based. For example, one graduate student found it easier to think about the number of particles than the intensity during the Intensity task:
\begin{center}
\begin{minipage}[c]{0.4\textwidth}
``I want to say the number of particles is proportional to the intensity\ldots I don't know if there's a square in there or not.''
\end{minipage}
\end{center}
They proceeded to solve the task relating number of particles to the depth of the liquid, and then translated the number of particles to intensity, and depth of the liquid to the total distance traveled by the probe. \\

\noindent
\textbf{Regions of Consistent Behavior} \\
The use of ``regions of consistent behavior'' is defined as dividing a task based on what parts of the domain make sense to be modeled by the same function. We observed that the boundaries between regions were often identified by an abrupt change in motion, and experts tended to focus on modeling one section at a time. MMA~2 was often used alongside regions of consistent behavior as considering whether a quantity was increasing or decreasing helped to identify possible boundaries between sections. All physics experts were observed to use regions of consistent behavior.

All of the participants divided Gravitation or Electric Charge (depending on which they saw) into sections. For example, one graduate student says upon watching the animation for Electric Charge:
\begin{center}
\begin{minipage}[c]{0.4\textwidth}
``Starting and finishing the same distance away, uh, getting closer and then being a constant distance away for a period of time. So it’s going to be three sections to my graph.''
\end{minipage}
\end{center}
Experts did not necessarily consider the sections ``in order.'' Rather, they were likely to model each section independently, in whichever order made the most sense to them. For example, many experts began Going Around Tacoma, Gravitation, or Electric Charge by first drawing the middle section.\\

\noindent
\textbf{Physically Significant Points} \\
The use of ``physically significant points'' is defined as choosing and plotting a small number points that hold physical meaning. For these tasks, this physical meaning might represent bounds of a quantity (the maximum or minimum) or inflection points (e.g., the sides of the Ferris wheel). We observed that experts often chose physically significant points as a first step towards modeling the tasks. Similar to regions of consistent behavior, choosing what makes a point significant often includes the use of MMA~2 alongside considering the physical context. All physics experts were observed to use physically significant points.

For example, one graduate student began Electric Charge by choosing some points:
\begin{center}
\begin{minipage}[c]{0.4\textwidth}
``So I start at some small positive, go to some bigger positive, stay there, and then go back to my initial point.''
\end{minipage}
\end{center}
\noindent
They plot four points to represent this story, and then complete their graph saying, ``And it'll go like one over $r$ in between.'' We note that this is distinct from the behavior of mathematics graduate students noted by Hobson and Moore---in that case, graduate students were observed to divide the task into equal intervals. In our case, all of the experts divided tasks not by equal intervals on the horizontal axis but by a physically meaningful moment in time (i.e. a sharp corner in the motion of the electric probe). \\

\noindent
\textbf{Neighborhood Analysis} \\
``Neighborhood analysis'' is defined by creating a smooth graph by drawing small line segments centered physically significant points that represent the slope of the graph. This was typically followed by connecting these small line segments with a smooth curve and a verbalization of the second derivative. Neighborhood analysis was used when participants did not have a ready-to-apply model from prior physics experience. It necessarily relies on MMA~4, as experts consider ``small chunks'' of change around a point. A majority of physics experts used neighborhood analysis.

For example, a graduate student checking the curve they drew for Drone stated:
\begin{center}
\begin{minipage}[c]{0.4\textwidth}
``Because like, the angle is changing really quickly near the middle. You kind of think the derivative should be bigger there. And it sort of is, like, if I look at two, you know, two points around the middle. Like it's changing a lot faster than say two points here [gestures to the far left of their graph].''
\end{minipage}
\end{center}
This graduate student observed that the angle appears to change faster at the apex of the curve than on the sides, and used that as justification for the slope tangent to the curve at corresponding points on the graph.\\

These reasoning devices often simplified the covariational reasoning tasks, reduced cognitive load, and reduced the amount of time experts directly compared the change of one quantity to that of another. However, we suggest that these devices appear to be in part habitual---they are likely a part of routine problem solving for these experts. Therefore, we suggest the relative lack of directly comparing change and rate of change
was likely a consequence of experienced problem solving rather than an intentional choice.

In addition to the devices described here, we also observed all of the 15 physics experts simplified tasks explicitly using symmetry and/or limiting cases.  In a recent study of expert problem solving techniques across scientific fields, 100\% of those interviewed were observed to consider what assumptions and simplifications they could make \cite{Price2021AAssessment}. This result adds to this and other research in physics education on simplification techniques and limiting cases \cite{White2023LimitingCourse}. Recent work has also reported that students who were prompted to evaluate the validity of proposed mathematical expressions for problems in introductory physics used a variety of strategies to do so including the canonically taught strategies of limiting/special cases, unit checking, and considering reasonableness of numbers \cite{AkinyemiSolutionStudents}.  As the reasoning devices we describe contribute novel research, and the observations we made about expert use of limiting cases and symmetry confirm previously published work, we focus our analysis in this report on novel characterizations of thought processes associated more closely with covariational reasoning.

\subsection{Modeling Modes}
\label{sec:resultsModes}
\begin{figure*}
    \centering
    \includegraphics[width=\textwidth]{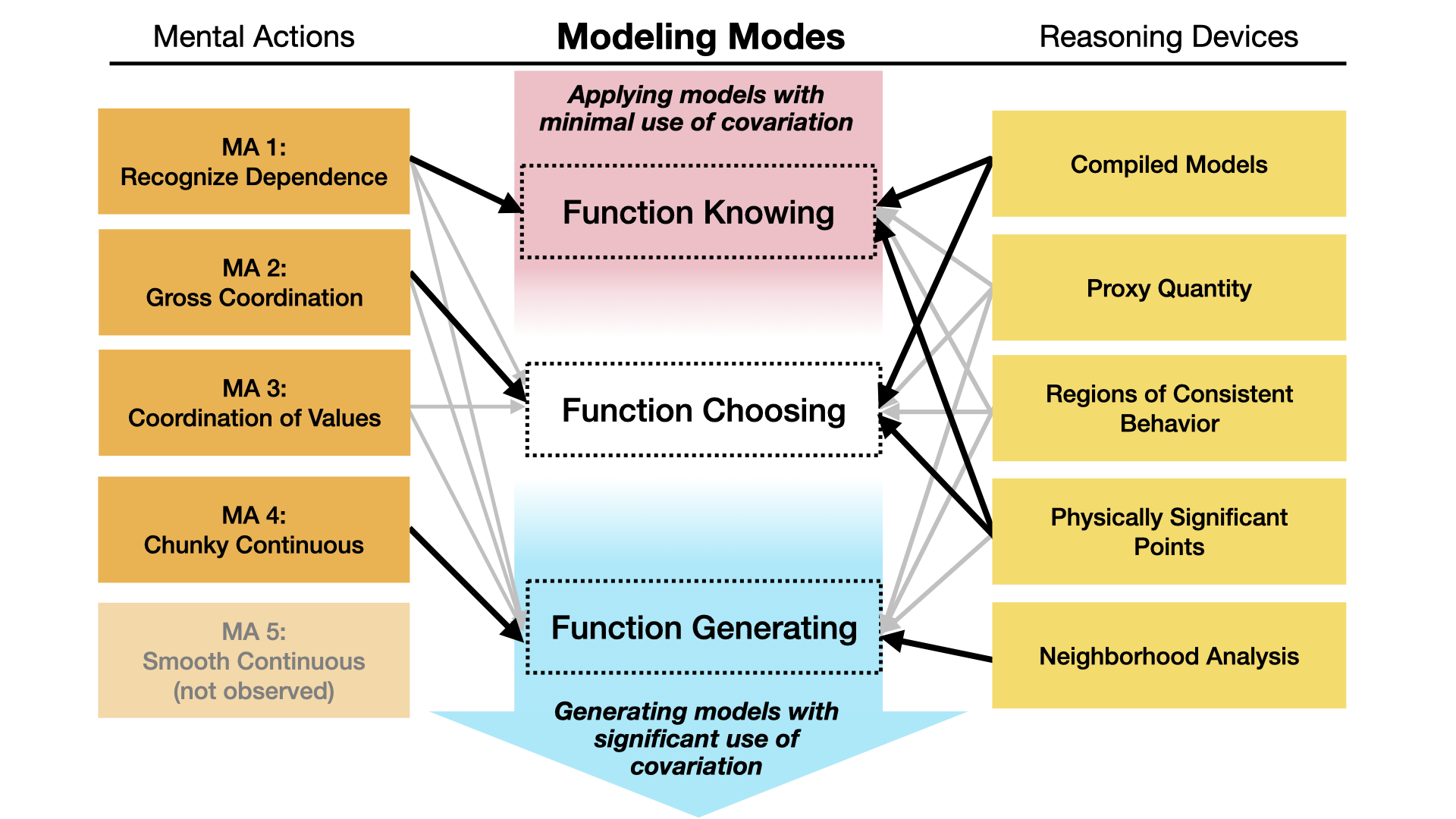}
    \caption{A description of the modeling modes we identified based on observations from the main study: Function Knowing, Function Choosing, and Function Generating. These modes are representative of a combination of mathematics mental actions (MMA, shown in orange on the left) and the reasoning devices we identified in Section~\ref{sec:resultsRD} (shown in yellow on the right). The black arrows represent the MMA and reasoning devices that are definitive of each mode, and the grey arrows represent MMA and reasoning devices that were common to a modeling mode, but not essential to its description. The central arrow represents the transition from applying models in Function Knowing to generating models in Function Generating.}
    \label{fig:modes}
\end{figure*}

Our second finding is that physics experts were likely to engage in one of three modes when graphically modeling the relationship between two quantities.  We call these modeling modes ``Function Knowing,''``Function Choosing,'' and ``Function Generating.'' They are larger patterns of reasoning that describe the integrated ways in which experts reasoned covariationally using a combination of the MMA and the reasoning devices (Tables~\ref{tab:StevensSummary} and~\ref{tab:RD}). This relationship is illustrated in
Fig.~\ref{fig:modes}. 
In this section, we describe how these modes were identified, how they are defined, and how they relate to one another.

During our examination of the participant timeline charts, we sought patterns between the code categories and in particular, those that included the MMA code category. A striking pattern emerged: there was a consistent interplay between the code categories ``Function'' and ``Mathematics Mental Actions'' (definitions can be found in Table~\ref{tab:maincategories}).
Nearly half of the time (48\%) that participant reasoning was assigned either code category, they were using one and then the other in quick succession. In addition, the code categories rarely appeared at the same time. The research team examined this pattern more closely by revisiting and discussing the original transcripts. We found that the data clustered to three modes that were defined by the individual codes used by experts within these categories. A more rigorous visualization of this analysis can be found in the Appendix.

In this paper, we use \textit{function} to refer to the functional form of a relationship between two quantities that describes how two quantities change together. This relationship can be expressed both graphically and symbolically. For example, a quadratic function can be represented symbolically $y \propto x^2$ and as a parabolic shape in which the dependent quantity varies smoothly in proportion to the independent quantity squared. For the purposes of this work, we consider both of these to be ``reasoning about function.'' Across tasks and expert populations, we observed physicists use the phrase ``goes as'' or ``goes like'' to refer to such a relationship (e.g., ``the height goes like a sine function''). We view the use of the phrase ``goes as'' or ``goes like'' to be a clue that someone is engaged in reasoning about a functional relationship between two quantities \cite{Zimmerman2020ExploringPhysics}.

\subsubsection{Defining Function Knowing, Choosing, and Generating}
In this section, we define Function Knowing, Choosing, and Generating and present evidence for how experts were observed to use them. We note that the distinctions between Function Knowing, Choosing, and Generating are not hard and fast; there were many instances where the research team held discussions about whether a particular portion of transcript was better characterized as one or another. We suggest that these modes describe a spectrum of approaches to modeling from recognizing a well-modeled physical situation towards generating a model where one does not exist. This is represented by the central arrow in Fig.~\ref{fig:modes}, which points from Applying Models to Generating Models.

\renewcommand{\arraystretch}{1.5}
\begin{table*}[]
    \centering
    \begin{tabular}{ p{0.25 \textwidth}  p{0.12 \textwidth} p{0.55 \textwidth}}
        \toprule
        Modeling Mode & Item & Example \\
        \toprule \vspace{-0.25cm}
        \multirow{2}{0.25\textwidth}{\textbf{Function Knowing} \qquad Quickly associating a physical context with a known mathematical model.}
            & Gravitation 
            & ``I want to talk about the gravitational potential energy of the entire system. Okay, so gravitational potential energy goes as $1/R$, one over the total radius to the system.'' 
        \\ \cmidrule{2-3}
            & Ferris Wheel
            & ``So, the total distance traveled is proportional to the time because it's moving at constant speed.''
        \\
        \toprule \vspace{-0.25cm}
        \multirow{2}{0.25\textwidth}{\textbf{Function Choosing} \qquad Plotting physically significant points and using the trend between them based on the physical context to choose a function.}
            & Intensity
            & ``And so we expect that at the maximum distance traveled, the intensity will be greatest. And as light is absorbed, it will get less great\ldots I have no idea. I would expect, and it depends if its in the linear regime, how exactly it will go. But naively, I'd expect that it'd be roughly linear.''
        \\ \cmidrule{2-3}
            & Intensity
            & ``So at the top, you'll get the maximum intensity here and it will drop---it will drop according to some some curve that depended on this. It could be an exponent---and likely an exponential. It really would---should be, uh---should be an exponential fall because the differential probability of having less light drives the thing.''
        \\
        \toprule \vspace{-0.25cm}
        \multirow{2}{0.25\textwidth}{\textbf{Function Generating} \qquad Connecting physically significant points with a smooth curve using the rate of change as a guide.}
            & Drone 
            &``The top---basically, the height changes the most slowly, and then the bottom\ldots and, well, as you increase the angles, the height goes---changes faster. So you should have a similar trajectory as you would see in the picture. But how exactly it is, is a different question. I think you can't solve this question quantitatively, but qualitatively you can draw a curve like this.''
        \\
        \bottomrule
    \end{tabular}
    \caption{The ways in which experts engaged with mathematical modeling while using covariational reasoning during graphing tasks.}
    \label{tab:mangoTangoExamples}
\end{table*}

\textbf{Function Knowing} is characterized by accessing a compiled model and applying it to the context using physically significant points (such as those described by the initial conditions of the task). Function Knowing often involves the use of proxy quantities to make sense of the accessed model in the context of the task. It is described by an interplay between compiled models, physically significant points, and MMA~1 (Fig.~\ref{fig:modes}). 

For example, upon seeing the Ferris Wheel task, many of our interviewees immediately identified the context as circular motion, and that they would need to use a trigonometric function (compiled models). Most then chose the phase based on the starting point of the cart (physically significant points) and reasoned about how to ``map'' between time and total distance to make sense of their initially accessed model that related height and time (proxy quantity, MMA~1). 

\textbf{Function Choosing} is characterized by using the general trend between physically significant points to choose between a handful of compiled models. It is described by an interplay between compiled models, physically significant points, and MMA~2 (Fig.~\ref{fig:modes}).

For example, during Intensity, many experts reasoned that the lowest intensity was at zero distance traveled and the highest at maximum distance traveled (physically significant points). Then interviewees chose how best to connect the points by using the general trend between them (MMA~2) and by reasoning about which functions made sense in the physical context---e.g. linear or exponential (compiled models).

\textbf{Function Generating} is characterized by using neighborhood analysis: determining the rate of change at a small number of physically significant points and then plotting the points along with small line segments that represent the slope of the curve at those points. The points are then connected with a smooth curve using the line segments as a guide. In 12 of the 18 instances we observed of function generating, participants named the function that looks like the curve they drew. In the rest, they simply ended the task after completing the graph. It is described by an interplay between physically significant points, neighborhood analysis, and MMA~4 (Fig.~\ref{fig:modes}).

For example, during Drone, some experts reasoned that the angle would change more quickly with respect to the height at the top of the arc than at the sides (physically significant points, MMA~4). One approach was to draw slopes to indicate these rates of change, and then connect them with a smooth line (neighborhood analysis). Instances when neighborhood analysis was used to verify or generate the graph were both considered Function Generating.

Examples of each pattern are shown in Table~\ref{tab:mangoTangoExamples}. These examples also help to demonstrate the distinctions between the three approaches. In the examples of Function Knowing, both speakers have a model ready at hand, and spend their time deciding how to apply it. In the examples of Function Choosing, the speakers use the general trend between points to choose between a handful of possible models. Finally, in the example for Function Generating, the speaker does not identify a possible function and instead is guided by their understanding of the first and second derivative. Function Generating was consistently observed to be a last-resort option: experts were likely to begin with Function Knowing or Choosing and finally engage in Function Generating only if Function Knowing or Choosing had failed. These approaches are emergent from the data and are therefore necessarily reliant on the ways the experts chose to talk about how they were reasoning. Despite this limitation, they give us insight into common approaches expert physicists took when solving graphical covariational reasoning tasks.

\subsubsection{Evidence of Function Knowing, Choosing and Generating}
We predominately observed Function Knowing and Choosing during tasks that were either constant motion contexts (i.e. Ferris Wheel) or tasks that required some physics knowledge to complete (i.e. Gravitation, Electric Charge, and Intensity). During Ferris Wheel, the majority of the participants (7 of the 10 graduate students and all of the faculty) held strong associations between the circular motion context and trigonometric functions. They quickly identified that they needed a sine function, and used initial conditions to identify the phase and the amplitude:
\begin{center}
\begin{minipage}[c]{0.4\textwidth}
``So our height is just going to be kind of going sinusoidal\dots [and the cart] started out at highest point [draws curve].''
\end{minipage}
\end{center}

Gravitation (and Electric Charge) and Intensity required physics knowledge to solve, and therefore necessarily required either Function Knowing or Choosing. However, these contexts were less familiar to most participants and we observed that graduate students were likely to toss out several different models while engaged in Function Choosing. For example, during the Intensity task, one graduate student stated:
\begin{center}
\begin{minipage}[c]{0.45\textwidth}
``Not going to use it, but intensity is proportional to the magnitude of the field squared.
\end{minipage}
\end{center}
\begin{center}
\begin{minipage}[c]{0.45\textwidth}
Yeah, I'm not going to use that. Because\dots no, I mean, never mind, I can't use that. That is what that's what intensity is. But I'm not going to calculate\dots
\end{minipage}
\end{center}
\begin{center}
\begin{minipage}[c]{0.45\textwidth}
Should I consider the situation\dots exponential from? I guess that's just me reaching for a model. That happens a lot. Exponential wouldn't make sense. That's like skin depth penetration. That would be like a beam hitting a conductor.''
\end{minipage}
\end{center}

This line of reasoning was ultimately productive for the graduate student, as they were able to get some initial compiled models out of the way by realizing that the models did not apply to the question being asked. However, they also distracted themselves from the correct answer (exponential) with other content knowledge.

This example is provided to show the range of Function Choosing---for some, it was a process of choosing between compiled models that had physical meaning in the context of the task. For others, it looked like coming up with several possible models from physics that are related to the quantities involved, and working through which were relevant to the prompt. Faculty were more likely to engage in the former approach, and early career graduate students were more likely to use the latter.

Function Generating was used primarily on the Drone task. This may reflect the fact that deriving an analytical solution is challenging and, as one faculty member said, ``messy.'' Nearly every expert solved Drone by engaging with Function Generating, either as a check on an initially drawn curve or to develop a curve.

\begin{figure}
    \centering
    \includegraphics[width=0.35\textwidth]{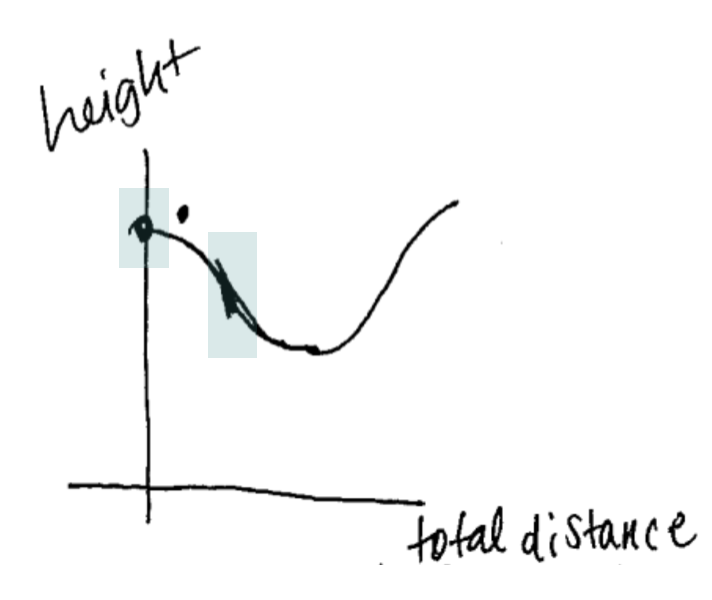}
    \caption{A graduate student's work from Ferris Wheel that demonstrates Function Generating reasoning; the point and line segment they used are highlighted in green.}
    \label{fig:fnGen}
\end{figure}

We also observed that the three graduate students who did not have a compiled model for Ferris Wheel (or who may have thought that they were not supposed to use one) used Function Generating to arrive at a graphical representation. These participants did not demonstrate an association between circular motion and trigonometric functions, and instead examined the rate of change at particular points in order to determine the graph. For example, one graduate student stated:
\begin{center}
\begin{minipage}[c]{0.4\textwidth}
``So those---kind of---be some kind of curvature to---to this.
\end{minipage}
\end{center}
\begin{center}
\begin{minipage}[c]{0.4\textwidth}
So towards the halfway point, when it's almost, when it's halfway to the bottom,
the speed should be the greatest. And then otherwise, it should be like---should kind of have a lower slope\dots
\end{minipage}
\end{center}
\begin{center}
\begin{minipage}[c]{0.4\textwidth}
So low to high slope back to low slope.''
\end{minipage}
\end{center}
\noindent
This student drew small line segments to guide their reasoning, and then connected them with a smooth line to form their final graph (Fig.~\ref{fig:fnGen}).

\begin{figure}
    \centering
    \includegraphics[width=0.5\textwidth]{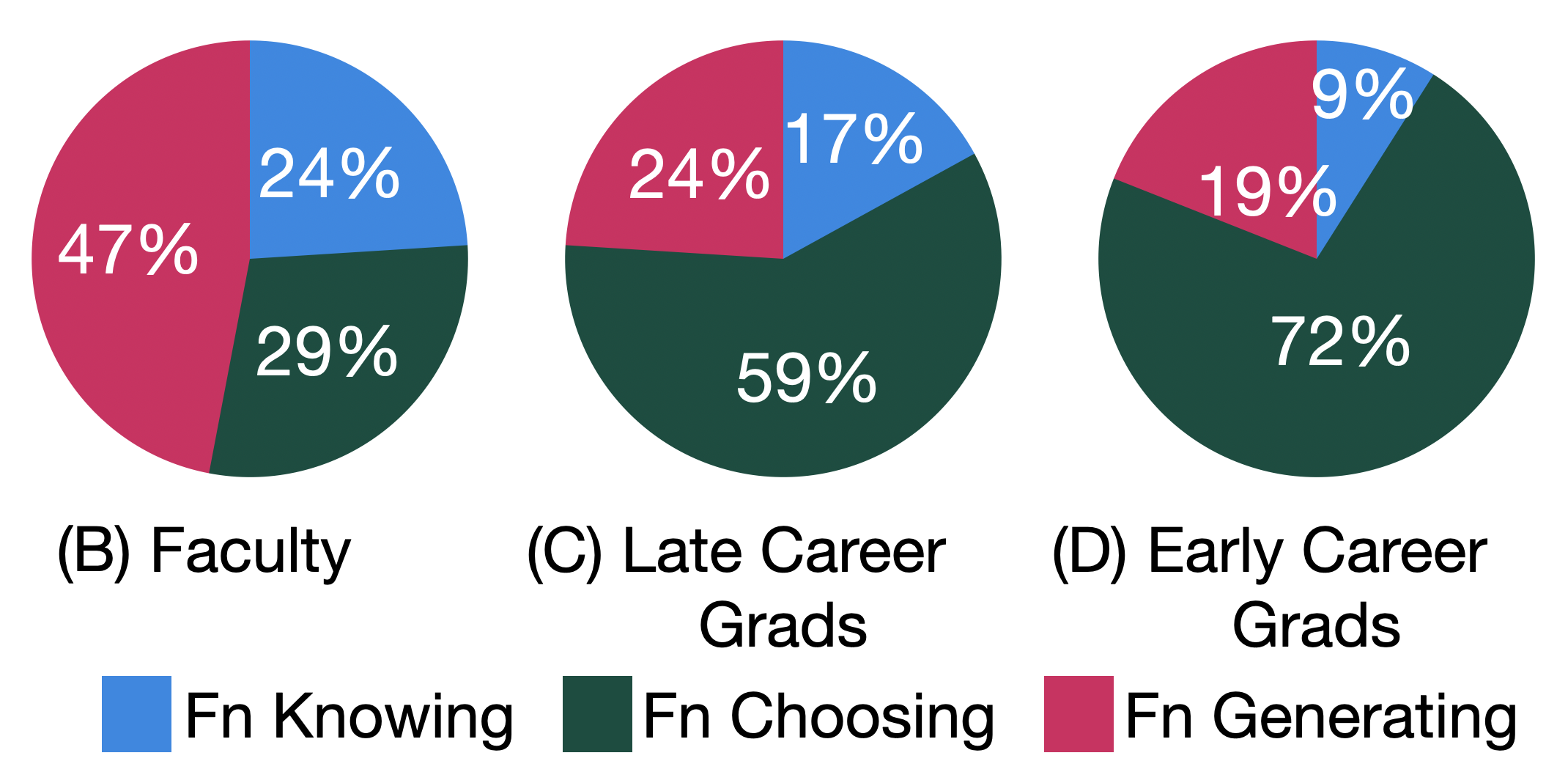}
    \caption{The time participants in Study 2 and 3 spent Function Knowing, Function Choosing, or Function Generating as a percentage of time spent engaged in function reasoning.}
    \label{fig:mangoTangoPieCharts}
\end{figure}

While all participants appeared to use all five reasoning devices, there were interesting differences in how the populations engaged in the modeling modes. We found the early career graduate students (those in their first year of study) were most likely to engage in Function Choosing, and faculty were most likely to engage in Function Knowing or Function Generating (see Fig.~\ref{fig:mangoTangoPieCharts}).

\section{Discussion}
\label{sec:discuss}
This work seeks to characterize covariational reasoning used by physics experts while solving novel graphing tasks. We found that some of the covariational reasoning mental actions defined by mathematics education research were consistent with physics experts' reasoning, and some were not. In addition, there were ways physics experts reasoned that were not well captured by the MMA. This prompted our research question, ``What features characterize the ways physics experts use covariation to solve novel graphing tasks?'' We identified reasoning devices and modeling modes that were used consistently by the majority of experts while reasoning covariationally. In this section, we describe how these findings relate to current literature and present a hypothesis for some aspects of physics covariational reasoning.

\subsection{The Emergent Reasoning Devices and their Relationship to the Mathematics Mental Actions}
We identified five reasoning devices used by experts to covary that are distinct but related to the mathematics mental actions for covariation.
We suggest these devices often engage MMA but include a blended consideration of the physical context. They therefore contribute to patterns of expert physics covariational reasoning when solving novel graphing tasks.

Proxy quantities were consistently used by physics experts across most of the tasks. In particular, we saw that proxy quantities allowed expert physicists to use models they already knew efficiently and ``map'' the familiar model to the contexts and quantities the task asked for. 
Proxy quantity necessarily relies on MMA~1, as a central behavior is relating quantities. However, essential to the distinction between proxy quantities and MMA~1 is the replacement behavior, which includes an assumption that one quantity is easier to interpret physically (or more common in accepted physical models) than another.

Use of proxy quantities is reminiscent of what mathematics education researchers have characterized as ``simultaneous covariation''  \cite{Johnson2015TogetherChange}: considering how two quantities change independently with respect to time and then comparing the independent changes of one to coincident changes in the other. However, while a small number of physics experts described their use of proxy quantities as ``parameterization,'' most simply talked about ``replacing'' distance with time, which we view as distinct. Further, we observed physics experts engaged in using proxy quantities that were not time, namely, radius for total distance in Gravitation and Electric Charge. 

We also observed experts sectioning tasks into regions of consistent behavior, and using neighborhood analysis to examine the rate of change at particular, physically significant points. When sectioning tasks, there was an interplay between MMA~2 (used to find points of inflection) and the physical interpretation of those inflection points. This interplay allowed experts to describe a region as ``consistent'' (i.e. articulating that one quantity increases while the other stays the same for a section of the domain), and is particular to physics covariational reasoning. A good example of this interplay is the Ferris Wheel task---while most experts considered the entire task one ``section'', 3 of the 15 separated the interval where the cart was going up from the interval when the cart was going down.

Similarly, we suggest that neighborhood analysis engages MMA~4, but in a different way than described by the existing frameworks. Thompson and Carlson's 2017 framework defines MMA~4 as comparing discrete change in two quantities; the next step (MMA~5) is smooth continuous variation, in which the person views the quantities changing smoothly together \cite{Thompson2017}. We suggest that neighborhood analysis sits between these two mental actions; physics experts had in mind that the quantities changed smoothly, but used small, nearly infinitesimal, chunks of discrete change modeled at particular points to guide their smooth curve.

\subsection{Modes Associated with Modeling using Physics Covariational Reasoning}
A broader pattern in how experts covaried while reasoning about function emerged from the data.
We observed three modes of approaches to reasoning covariationally, both symbolically and graphically, which we call Function Knowing, Function Choosing, and Function Generating. In this section, we propose a hypothesis for how these approaches are used and describe how they represent some important ways that expert physicists reason covariationally. 

We observed that experts were likely to begin with Function Knowing (if accessible), then work through Function Choosing, if more than one compiled model surfaced. Function Generating was consistently the last-resort option, and only attempted once Function Knowing or Choosing was eliminated. From our findings, we described the arrow in Fig.~\ref{fig:modes} as descriptive of the observed reasoning progression.

In our model, Function Knowing, Choosing, and Generating are characterized by the combined use of MMA and the reasoning devices described in Table~\ref{tab:RD}. 
Function Choosing, Knowing, and Generating are all founded on reasoning about quantities in the context of the physical world and their change with respect to one another. Thus, we propose that Function Choosing, Knowing, and Generating are three patterns associated with expert physics covariational reasoning. 

\subsection{Comparing Faculty and Early Career Graduates}

 We observed that early career graduate students spent more time trying out different functions (Function Choosing), and faculty were more comfortable generating functions from scratch---they were more likely to turn to Function Generating when they realized they did not have a ready-to-use model to apply (see Fig.~\ref{fig:mangoTangoPieCharts}). We view this distinction as characteristic of what Bing and Redish describe as a \textit{journeyman} level in physics that is situated between expert and novice. Journeymen are characterized as ``having sufficient skills that they can no longer be considered novices, but not yet had sufficient experience with sophisticated problem solving (and research) to be considered experts'' \cite{Bing2012EpistemicTransition}. 
 For early-career graduate students still taking courses, mathematically complex problems with closed-form solutions are the norm. It may be that students at this stage in their careers are less likely to solve the tasks we provided with a generative frame of mind. They may have an expectation of a ``right answer'' that can be expressed as a single function, and we observed many searching for that in earnest. 
 We interpret our results as reflecting the relative lack of experience in research and 
 ``test-taking'' mindset of early graduate school. Our data may imply that by engaging students in the generative aspects of research as part of their coursework, graduate programs can provide opportunities for students to hone the important expert practices of modeling quickly and making simplifying assumptions. Our interpretations are preliminary; more research is needed to fully investigate these ideas.

\section{Conclusion}
\label{sec:conclusion}

Covariational reasoning has been defined by mathematics education research as ``the cognitive activities involved in coordinating two varying quantities while attending to the ways in which they change in relation to each other'' \cite{Carlson2002ApplyingStudy}. Our empirical results show that the mental actions that resulted from this definition are productive in physics education research, and that there are a number of ways physicists reason covariationally that are not included in those mental actions. We identify five reasoning devices distinct from the mathematics mental actions used by physics experts to solve graphing tasks. These devices are used in conjunction with the mathematics mental actions to form patterns that we identify as modeling modes.

We recognize that this work is a first step towards identifying expert physics covariational reasoning. As a qualitative, generative experiment, there is significant opportunity to revisit these results with additional experts and coders. Future work includes a hypothesis testing experiment in which our codebook could be used by a larger group of researchers and applied to a wider pool of experts. There is also a need to expand beyond the graphing tasks we have shown here to explore how physics experts engage with more symbolic-based covariational reasoning. 

Another essential next step of this work is to interview mathematics faculty on their covariational reasoning. This study was designed to investigate the ways in which physics experts used covariation that goes beyond the mathematics mental actions. It did not include a comparison to identical populations with the same tasks to probe whether these reasoning devices are also common in mathematics. One limitation to such a study is the level of physics content knowledge required to answer the tasks we designed. One would likely need a new set of tasks that do not require physics content knowledge but do go beyond familiar models for physicists.

We suggest that these reasoning devices and modeling modes can also provide a platform for instructors to reflect on their own reasoning. Understanding expertlike reasoning devices can help instructors support physics students in developing expertise. For example, plotting points is a behavior that many students may view as overly simple or a last resort. Helping students to recognize that identifying and plotting specific, physically meaningful points is an expertlike tool for problem solving is just one instructional application of this work. In addition, there is an opportunity to design activities that help students learn to graph in a generative way to make sense of how two quantities are related. 

The work described in this paper provides an empirical foundation for operationalizing physics covariational reasoning. As a next step, we are 
working to characterize the understanding of the mathematical foundations and physics quantities that seem to be necessary for expertlike physics covariational reasoning. These findings will be synthesized into a framework of covariational reasoning in physics (CoRP). Our hope is that the CoRP framework will not only operationalize physics covariational reasoning but will also be productive for characterizing physics students' covariational reasoning.

\section{Acknowledgements}
We would like to thank the faculty and students at the University of Washington who have made this work possible---their participation and support has been integral to this project. We would also like to thank Peter Shaffer and John Goldak for their thoughtful reviews during the preparation of this material. This work was partially supported by the National Science Foundation under grants No. DUE-1832836, DUE- 1832880, DUE-1833050, DGE-1762114, DUE-2214765, and DUE-2214283. Some of the work described in this paper was performed while the second author held an NRC Research Associateship award at Air Force Research Laboratory.

\bibliography{references}

\appendix*
\section{Preliminary Codebook}

\renewcommand{\arraystretch}{1.1}
\begin{table*}
\begin{tabular}{p{0.15\textwidth} p{0.22\textwidth} p{0.6\textwidth}}
Code Category & Code & Description \\
\hline   
    & Identifying Quantity 
        & Identifying a quantity given in the problem, i.e. ``The distance from Tacoma is\ldots''. \\
\multirow{5}{0.15\textwidth}{Considering Quantity}
    & Defining Quantity 
        & Defining a new quantity, i.e. labeling a radius not stated in the prompt.\\
    & Classifying Quantity 
        & Connecting a quantity with its type or kind (i.e. vector or scalar). \\
    & Representing Quantity 
        & Choosing or expanding upon a representation of a quantity, i.e. defining axes labels or variables.\\
    & \cellcolor{Apricot}  Relating Quantity
        & \cellcolor{Apricot} Explicit discussion of how / if one quantity depends on another. \\
    & \cellcolor{Apricot} 
        & \cellcolor{Apricot} \emph{*This code was identified as MMA~1 after analysis was complete.}\\
\hline 
    & Image of Change 
        & Referring to change as discrete or continuous. \\
\multirow{5}{0.15\textwidth}{Approach to the Problem}
    & \cellcolor{Goldenrod} Physically Significant Points 
        & \cellcolor{Goldenrod} Identifying a specific point or set of points that are important to solving the problem using the representation or problem statement given. \\
    & \cellcolor{Goldenrod} Regions of Consistent Behavior 
        & \cellcolor{Goldenrod} Identifying a section of the representation or graph that has distinct or unique behavior; the process of sectioning off regions of the problem to focus on individually. \\
    & Visualizing 
        & Attempting to ``see'' what’s happening as a way to explain their reasoning. \\
    & Axes Preference 
        & Articulating a desire to put certain quantities on certain axes. \\
\hline
\multirow{4}{0.15\textwidth}{Mapping Math to Physics}
    & \cellcolor{Apricot} Increase/Decrease
        & \cellcolor{Apricot} Using language of increases or decreases to define the behavior of a quantity with respect to another.\\
    & \cellcolor{Apricot} 
        & \cellcolor{Apricot} \emph{*This code was identified as MMA~2 after analysis was complete.}\\
    & \cellcolor{Goldenrod} Determining Function
        & \cellcolor{Goldenrod} Attempts to link a kind of function to a section of a graph. \\
    & \cellcolor{Goldenrod} Assigning Function
        & \cellcolor{Goldenrod} Deciding on a particular mathematical function for a chosen part of a graph. \\
\hline
\multirow{5}{0.15\textwidth}{Trends of Change in Quantity}
    & \cellcolor{Apricot} Rate of Change 
        & \cellcolor{Apricot} Discussing a rate of change (a change of one quantity with respect to another) at a particular point or across a region.\\
    & \cellcolor{Apricot} 
        & \cellcolor{Apricot} \emph{*This code was identified as MMA~3, 4 after analysis was complete.}\\
    & \cellcolor{Goldenrod} Neighborhood Analysis
        & \cellcolor{Goldenrod} Examining the slope or changing behavior of a quantity around a specific point; in its ``neighborhood'' \\
    & Comparing Rates
        & Explicitly discussing how the rate of change at one point in time compares to another; can be between parts and tasks \\
    & Comparing Slopes
        & Explicitly discussing how the slope of one part of the graph relates to another; can be between parts and tasks. \\
\hline
\multirow{3}{0.15\textwidth}{Compiled Models}
    &  \cellcolor{Goldenrod} Producing a Model 
        &  \cellcolor{Goldenrod} Developing a separate relationship between two quantities that describes one part or a simplification of the task at hand.\\
    &  \cellcolor{Goldenrod} Applying a Model
        &  \cellcolor{Goldenrod} Explicit reference to a model they produced or have prior knowledge of to address the task at hand. \\
    &  \cellcolor{Goldenrod} Eliminating a Model
        &  \cellcolor{Goldenrod} Deciding that a model doesn’t apply to the current situation.\\
\hline
\multirow{3}{0.15\textwidth}{Graphical Structure}
    & \cellcolor{Goldenrod} Connecting Slopes 
        & \cellcolor{Goldenrod} Using known slopes at points to create a smooth line in a graph \\
    & Generating Graphical Features
        & Using the representation or problem statement to identify what graphical features must be present. \\
    & Validating Graphical Features 
        & Using a graphical feature to check expected behavior. \\
\hline
\multirow{3}{0.15\textwidth}{Simplification by Comparison}
    & Comparing Parts
        & Relating sections of representations or graphs to each other. \\
    & Comparing Tasks
        & Relating representations or graphs of separate tasks to each other.\\
\hline 
\multirow{3}{0.15\textwidth}{Sensemaking}
    & Real World Conflicts
        & Having to reconcile their lived experience and the problem statement. \\
    & Real World Informs 
        & Using their lived experience to aid in solving the problem. \\
    & Connecting Physics 
        & Connects the idea they’re working on to a larger concept or other concepts in physics \\
\hline
\end{tabular}
    \caption{The codebook from the Preliminary study. Codes that were interpreted during analysis to overlap with mathematics covariational reasoning mental actions are highlighted in orange. Codes that appeared unique when compared to that reported by Hobson and Moore are highlighted in yellow \cite{Hobson2017}.}
    \label{tab:codebookprelim}
\end{table*}

\section{Main Study Codebook}

\begin{table*}
\begin{tabular}{p{0.15\textwidth} p{0.22\textwidth} p{0.6\textwidth}}
Code Category & Code & Description \\
\hline   
\multirow{5}{0.15\textwidth}{Quantity}
    & Symbolizing
        & Representing or defining quantity with symbols, i.e. drawing a picture, identifying variables.\\
    & Constructing Quantity 
        & Defining a quantity by others, or making sense of why a quantity is constructed of others. \\
    & Mathematical Structure
        & Choosing a particular structure or discussing why a structure makes sense for a particular quantity, i.e. whether it can be negative, scalar, etc. \\
    & \cellcolor{Goldenrod} Proxy Quantity
        & \cellcolor{Goldenrod} Using one quantity in place of another to make sense of what is happening or visualize. \\
    & Recognizing Quantity
        & Choosing and naming (could include defining) which quantities are important in a given problem.\\
\hline
\multirow{5}{0.15\textwidth}{Mathematics Mental Actions (MMA)}
    & \cellcolor{Apricot} MMA~1
        & \cellcolor{Apricot} Relating quantity\\
    & \cellcolor{Apricot} MMA~2
        & \cellcolor{Apricot} Trend of change \\
    & \cellcolor{Apricot} MMA~3   
        & \cellcolor{Apricot} Discrete change \\
    & \cellcolor{Apricot} MMA~4
        & \cellcolor{Apricot} Small chunks of change \\
    & \cellcolor{Apricot} MMA~5
        & \cellcolor{Apricot} Smooth continuous \\
\hline
\multirow{4}{0.15\textwidth}{Function}
    & Expecting Simplicity
        & Expressing an epistomological perspective that there will be sufficiently described by a parent function.\\
    & Expecting a Solution
        & Expressing an epistomological perspective that there will be an analytical solution.\\
    & \cellcolor{Goldenrod} Compiled Models
        & \cellcolor{Goldenrod} Identifying that a quantity is proportional, or goes like, a particular function or other quantity; can also include assigning a common, simple function to a particular context (linear, quadratic, inverse, etc.)\\
    & Constant
        & Identifying a constant quantity or that a relationship between quantities is constant.\\
\hline
\multirow{6}{0.15\textwidth}{Graphing}
    & Constructing a Graph
        & Behaviors associated with actually drawing a graph (i.e. labeling axis, drawing a line, etc.) \\
    & Choosing (In)dependent Variables
        & Making a choice about what quantities make sense to be dependent on others, or if a quantity is independent. This can include making choices about which makes more sense to be on the horizontal axis. \\
    & \cellcolor{Goldenrod} Physically Significant Points
        & \cellcolor{Goldenrod} Choosing to focus particularly on points that have important, physical meaning. \\
    & \cellcolor{Goldenrod} Connecting Points
        & \cellcolor{Goldenrod} Choosing to draw a graph by connecting up points in a particular way (i.e. no particular function is expresssed as being in mind).\\
    & Sketching to Think
        & Expressing an understanding of what a quantity will or should look like; graphing as a part of the quanitification process.\\
    & Smooth
        & Expressing that it is both true and justified that most physical behaviors are smooth.\\
\hline
\multirow{5}{0.15\textwidth}{Simplification Techniques}
    & Symmetry
        & Using symmetry to simplify or make sense of a graph.\\
    & Limiting Cases
        & Examining limits and limiting cases of a physical situation, graph, or quantity.\\
    & Simplifying Assumptions
        & Making assumptions to simplify the problem (i.e. assuming constant speed, linear trends, etc).\\
\hline
\multirow{1}{0.15\textwidth}{Connection to the Task}
    & Connection to the physical world
        & Making sense of or discussing a quantity's physical meaning in the context of the task; drawing a direct connection between what is observed in the task and a representation. \\
\hline
\end{tabular}
    \caption{The codebook from the Main study. Codes that were interpreted during analysis to overlap with mathematics covariational reasoning mental actions are highlighted in orange. Codes that appeared unique, and were named as reasoning devices, are highlighted in yellow.}
    \label{tab:codebookmain}
\end{table*}

\section{Function and Mental Actions Timeline Charts}

\begin{figure*}
    \centering
    \includegraphics[width=\linewidth]{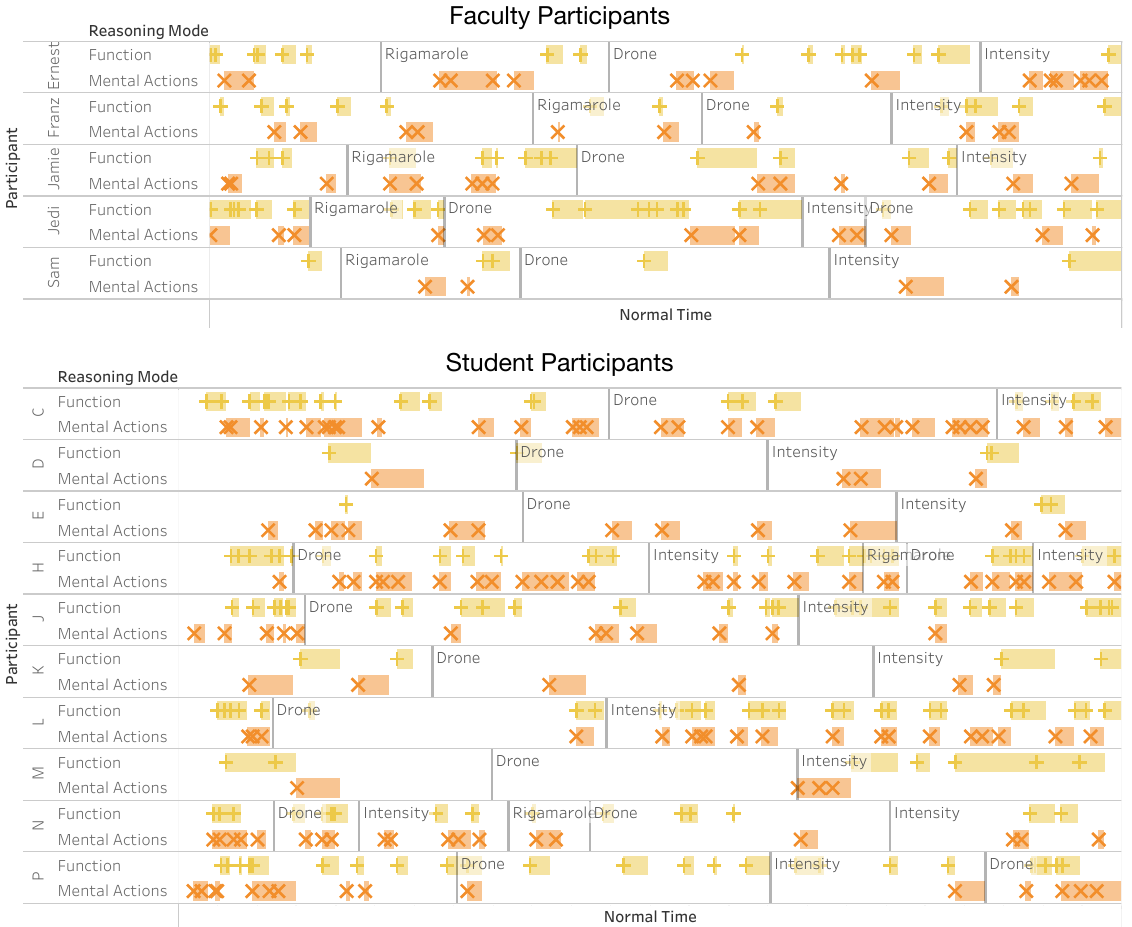}
    \caption{The timeline charts for the Function and Mental Actions code categories for all participants. Function Knowing, Choosing, and Generating are defined by the moments in which participants ``go between'' using function-like reasoning and the mathematical mental actions. This appears in the charts as times when participants are coded with either yellow or orange, and then in the next instant are coded with the other color. Strong examples of this behavior can be seen in Student C's timeline during the first portion of their interview, and when faculty member Jamie is reasoning about the Rigamarole task.}
    \label{fig:mangoTangoTimelines}
\end{figure*}

\end{document}